\documentclass[12pt,a4paper]{article} 
\usepackage{epsfig}
\usepackage{graphics}
\usepackage{subfigure}
\usepackage{a4wide}
\usepackage{amsmath}

\newcommand{\ie}{{\it ie.}}
\newcommand{\cf}{{\it cf.}}
\newcommand{\half}{{\scriptstyle\frac{1}{2}}}
\newcommand{\sech}{\mathop{\rm sech}\nolimits}

\newcommand{\RR}{{\bf R}}
\newcommand{\cL}{{\cal L}}
\newcommand{\cM}{{\cal M}}
\newcommand{\vecp}{\vec\phi}
\newcommand{\pa}{\partial}
\newcommand{\zb}{\overline{\zeta}}
\newcommand{\ee}{{\rm e}}
\newcommand{\ii}{{\rm i}}

\renewcommand{\a}{\alpha}
\renewcommand{\b}{\beta}
\newcommand{\m}{\mu}
\newcommand{\n}{\nu}
\newcommand{\vp}{\varphi}
\newcommand{\s}{\sigma}
\renewcommand{\S}{\Sigma}
\newcommand{\g}{\gamma}
\renewcommand{\l}{\lambda}
\renewcommand{\O}{\Omega}
\newcommand{\z}{\zeta}

\title{Stability of Sigma-Model Strings and Textures.}
 \author{R S Ward\footnote{email: richard.ward@durham.ac.uk}
 \bigskip
\\Department of Mathematical Sciences,  \\ University of
Durham, \\Durham DH1 3LE}

\begin{document}
\maketitle \abstract{\noindent
In flat space-time, sigma-model strings and textures are both unstable
to collapse and subsequent decay.  With sufficient cosmological expansion,
however, they are stable in a generalized sense: a small perturbation
will cause them to change their shape, but they do not decay.  The current
rate of expansion is sufficient to stabilize strings, but not textures.
}

\section{Introduction}

This letter deals with topological-soliton solutions of the O($n$)
sigma-model in curved space-time, and in particular with the effect
of cosmological expansion.  For topological defects with a fixed
length-scale (such as abelian-Higgs strings), the effect of
expansion on an individual soliton is not very significant (of course it
affects the overall density of solitons, but that is not the
concern here).  Sigma-model solitons, however, do not have a fixed
size --- and so expansion can have a large effect (in particular, the
conclusions of Derrick's theorem can be circumvented).

We concentrate here on three kinds of soliton, namely textures in the O(4)
and O(3) systems, and strings in the O(3) system.  So the relevant
homotopy groups are $\pi_3(S^3)$, $\pi_3(S^2)$ and $\pi_2(S^2)$
respectively.  In flat space-time, any texture configuration will
immediately collapse; but there exist static string solutions.  The latter
correspond to solutions on $\RR^2$ (or more generally on a 2-surface $\S$),
which one may interpret as static straight strings extended in the
$z$-direction.  The (2-dimensional) energy of the configuration is
bounded below by $4\pi|N|$, where $N$ is the winding number; and one can
explicitly write down the solutions which saturate this bound \cite{BP75}.
In fact, one can also include the back-reaction on the
space-time metric, obtaining the corresponding solution of the
Einstein-sigma equations; and this solution saturates a lower
bound on the $C$-energy \cite{CG88}.  But the fact that the energy
is at its topological minimum does not mean that the solution is stable.
This is because the
scale of the soliton is not fixed --- in particular, the width of the
string is not fixed.  So a small perturbation may cause the string to
spread out in space, or to become infinitesimally thin and decay.
(Decay occurs when field gradients become large, so that the field $\vec\phi$
is able to climb out of its potential well where $\vec\phi^2 = {\rm const}$).
The only sure way to prevent the string from collapsing is to add a
`hard core' with a fixed size --- this is, for example, what happens for
semilocal strings \cite{VA91,H92,L92} where the core is an abelian-Higgs
string.

In what follows below we shall see that there are exact comoving string
solutions in an expanding universe, and that they are stabilized
by the expansion.  For textures, collapse could also be prevented by
expansion, but the required rate of expansion is higher than that in
the post-inflationary era.  In all cases, the stability is of a generalized
kind: a perturbation will cause solitons to change their shape (or size)
permanently.  Roughly speaking, the solitons are plastic rather than elastic.

%%%%%%%%%%%%%%%%%%%%%%%%%%%%%%%%%%%%%%%%%%%%%%%%%%%%%%%%%%%%%%%%%%

\section{Comoving String Solutions.}

The O(3) sigma-model involves a unit-vector field
$\vecp=(\phi^1,\phi^2,\phi^3)$ depending on the space-time coordinates
$x^\m$, with dynamics determined by the Lagrangian density
$\cL=\half g^{\m\n}(\pa_\m\vecp)\cdot(\pa_\n\vecp)$.  In what follows,
we shall also make use of the stereographic
projection $W = (\phi^1 + \ii\phi^2)/(1-\phi^3)$.

On a two-dimensional space $\S$, the system is conformally invariant.
Let us take $\S$ to be either $\RR^2$ (in which case we impose the
boundary condition $\phi^3\to1$ as $r\to\infty$) or $S^2$.
If we choose local complex coordinates $(\z,\zb)$ on $\S$ such
that its metric is $d\s^2 = \O(\z,\zb)^2\,d\z\,d\zb$, then any meromorphic
function $W(z)$ is a solution of the field equations on $\S$; and in fact
all solutions on $\S$ are of this form.  The lower bound  $E\geq4\pi N$ on
the energy of configurations is saturated if and only if $W$
is a rational function of $\z$, of degree $N$.

For example, $W(\z)=\l \z$ denotes a 1-soliton, located at $\z=0$, with
arbitrary size $\l^{-1}>0$.   So there is a potential instability: the
soliton may, for example, shrink to zero size (and become singular) in
finite time.  In flat (2+1)-dimensional space-time, this is exactly what
happens \cite{LPZ90,PZ96,BCT01,LS01}.  So a straight sigma-model string
in Minkowski space-time is unstable to cylindrically-symmetric perturbations:
such a perturbation can cause its width to shrink to zero in finite time.

The static solution of the full coupled system was described in
\cite{CG88}.  The energy-momentum tensor is
$T_{\m\n} = \kappa[(\vec\phi_{\m}\cdot\vec\phi_{\n}
            -\half g_{\m\n}(\vec\phi_{\a})^2]$,
the sigma-field is $W(\z)=\l \z$ as before, and the metric is
\[
  ds^2 = dt^2 - dz^2 - \frac{d\z\,d\zb}{(1+\l^2|\z|^2)^{16\pi G\kappa}} \,.
\]
For $0\leq16\pi G\kappa\leq1$, the surface $\S$ orthogonal to the string is
asymptotically conical, with a fixed defect angle; the apex of the cone
is smoothed out, but becomes non-smooth in the zero-width limit $\l\to\infty$.
Moduli-space approximations \cite{SS99,G01} suggest that the string in
the coupled system is unstable just as in the flat ($\kappa=0$) case,
but this is not known for certain.

These solutions can easily be generalized to an expanding universe.
For example, in the $\kappa=0$ case (no back-reaction), we can
take as background space-time the flat-space Robertson-Walker universe
\begin{equation} \label{RW}
 ds^2 = dt^2 - A(t)^2\,[d\z\,d\zb + dz^2];
\end{equation}
and then $W(\z)=\l \z$ is a solution representing a comoving string.
In the $\kappa\neq0$ case, $W(\z)=\l \z$ together with the metric
\[
  ds^2 = dt^2 - \ee^{2at}\left[dz^2 -
    \frac{d\z\,d\zb}{(1+\l^2|\z|^2)^{16\pi G\kappa}}\right]
\]
is a solution of the Einstein-sigma equations with
cosmological constant $\Lambda=3a^2$.

The moduli-space approximation referred to above consists of mechanics
on the finite-dimensional moduli space $\cM_N$ of static $N$-soliton solutions.
In other words, we consider only a finite number of degrees of freedom,
by restricting to the finite-dimensional subspace $\cM_N$ of the full
(infinite-dimensional) configuration space of the system.
For example, this was proposed for BPS
monopoles \cite{M82}, and in that case was shown to provide an accurate
approximation \cite{Stu94}.  It has been applied to several other systems,
and in particular to the sigma-model \cite{W85,L90,LPZ90,Sp95,PZ96,SS99,G01}.
In our case, the (inverse) width parameter $\l$ is one of the moduli,
and so we may consider the dynamics of $\l$.  The approximation is a useful
one; it breaks down when radiation becomes significant, and this happens if
the soliton collapses to zero width \cite{BCT01}.

In this case, the reduced Lagrangian $L$ for dynamics on moduli
space consists of a `kinetic energy' part only --- the `potential energy'
is equal to its constant topological value ($4\pi N$ above), and so is
irrelevant.  This kinetic energy involves an integral over $\S$, which in
the case $\S=\RR^2$ may diverge \cite{W85}.
One way of dealing with this is to restrict the integral to a compact
subregion of $\RR^2$, such as a disc $D_R$ of radius $R$, where $R$ 
is larger than the timescale of the process being studied \cite{PZ96,LS01}.
Another way is to work on $S^2$, where compactness ensures convergence, and
so the
reduced Langrangian $L$ is well-defined.  It has been has been calculated
explicitly in the $N=1$ case \cite{Sp95}; in particular, the
Langrangian for $\l$ is as follows.  We use a parameter
$\m$ related to $\l$ by $\l=(\sqrt{1+\m^2}+\m)^2$.  The
reduced Lagrangian is $L(\dot\m,\m)=h(\m)\,\dot\m^2$, where
\begin{equation} \label{h}
  h(\m) = \frac{8\pi\l^2\left[(\l^2+1)\log\l^2-2\l^2+2\right]}
               {(1+\m^2)(\l^2-1)^3} \,.
\end{equation}
Note that $\m\to\infty \Leftrightarrow \l\to\infty$ corresponds
to the soliton shrinking to zero size, while $\m=0\Leftrightarrow \l=1$
corresponds to the soliton being spread
out homogeneously over $S^2$.  In effect, $\m$ is a dimensionless
quantity which represents the ratio (size of $S^2$)/(size of soliton);
so we can model the case $\S=\RR^2$ (rather than $\S=S^2$) by taking
$\m\gg1 \Leftrightarrow \l\gg1$.

%%%%%%%%%%%%%%%%%%%%%%%%%%%%%%%%%%%%%%%%%%%%%%%%%%%%%%%%%%%%%%%%%%

\section{Stability of Strings.}

The aim here is to investigate the stability of sigma-model
strings in a fixed background space-time.  It has already been
remarked that in flat space-time, the string is unstable under
cylindrically-symmetric perturbations.  This can easily be seen
in the $\l$-approximation, as follows.  Take $\S$ to be a 2-sphere of radius
$R\gg1$, and take $\l\gg1$; so the effective Lagrangian becomes
$L\approx4\pi R^2\l^{-4}\log(\l)\,\dot\l^2$.  The corresponding
Euler-Lagrange equation can be reduced to the quadrature
\[
  \int \frac{\sqrt{\log\l}}{\l^2}\,d\l = Kt\,.
\]
So if $\l(0)=\l_0$ and $\dot\l(0)=\dot\l_0$ are the initial values,
then the time $T$ that it takes for $\l(t)$ to reach $\infty$ (\ie\ for the
string width to shrink to zero) is finite: in fact, $T\approx\l_0/\dot\l_0$.
Comparison with numerical solution of the full system (for example,
\cite{PZ96}) shows that even this rough approximation is a reasonable one,
underestimating $T$ by around $15\%$.

In an expanding universe, however, the string width does not shrink to zero
(if the perturbation is small). For the $\l$-approximation, the
argument is as follows.  In a Robertson-Walker background,
the effective Lagrangian is $L(\dot\m,\m) = A(t)^3\, h(\m)\,\dot\m^2$,
and the corresponding equation of motion can be reduced to
\[
  \int \sqrt{h(\m)}\,d\m = K \int A(t)^{-3}\,dt.
\]
It follows that if $T := \int_{t_0}^{\infty} A(t)^{-3}\,dt$
converges, and if the initial values $\m_0$ and $\dot\m_0$ satisfy
\[
 A(t_0)^3\,\sqrt{h(\m_0)}\,T\,\dot\m_0 \leq
        \int_{\m_0}^{\infty} \sqrt{h(\m)}\,d\m
\]
(which will always be the case if $\dot\m_0$ is small enough),
then the string is stable, at least in a generalized sense.  Namely, its
width will not shrink indefinitely as $t\to\infty$, but will settle
down to a value different from the initial one.  The perturbation causes
the string to shrink a bit, but then the cosmological expansion
stabilizes it and it shrinks no further.  So an expansion factor
$A(t)\sim t^p$ with $p>1/3$ should ensure stability in this sense.

This conclusion was checked by numerical solution of the
cylindrically-symmetric equation of motion, taking $\vecp$ to have the
form
\[
  \vecp = (\sin{f}\cos{\theta},\sin{f}\sin{\theta},\cos{f}),
\]
where $r$ and $\theta$ are polar coordinates on $\S=\RR^2$,
and where $f=f(t,r)$.  The equation for $f$ is
\begin{equation} \label{rot-sym-eqn}
 \frac{1}{A}\frac{\pa}{\pa t}\left(A^3\frac{\pa f}{\pa t}\right)
   = \frac{1}{r}\frac{\pa}{\pa r}\left(r\frac{\pa f}{\pa r}\right)
     - \frac{\sin{(f)} \cos{(f)}}{r^2}\,,
\end{equation}
which admits the comoving solution $f(r) = 2\cot^{-1}(\l r)$. For example,
taking $A(t)=t^{2/3}$ and initial conditions (at $t=1$) corresponding to
$\l_0 = 1$, $\dot\l_0=0.2$ gives a solution in which the width of the
string initially shrinks, but the shrinking levels off and stabilizes.
The function $f(t,r)$ as $t\to\infty$ is very close to
$f(r) = 2\cot^{-1}(\l r)$ with $\l=1.24$.

We can also consider perturbations which are localized in $z$
(rotationally- but not cylindrically-symmetric).  This was done using an
approximation in which the  field is $W=\l\z$
with $\l=\l(t,z)$.  The effective Lagrangian is
\[
   L = A(t)\,h(\m)\,[A(t)^2\,\m_t^2 - \m_z^2 ]\,,
\]
where $\m$ and $h(\m)$ are as defined previously.  The corresponding
Euler-Lagrange equations were solved numerically, with initial conditions
$\m(1,z)=10$, $\m_t(1,z)=\sech{(z)}$.  This represents making a constriction
in the string, localized around $z=0$.  
If $A(t)\equiv0$ (flat space-time), then the `dent' spreads outwards from
$z=0$ (at the speed of light), with $\m(t,z)$ staying finite everywhere.
With $A(t)=t^{2/3}$, however, the spreading-out is strongly damped:
the dent remains well-localized around $z=0$, and its maximum depth
(which is related to $\m(t,0)$) tends to a finite value as $t\to\infty$.
Once again, we see that cosmological expansion stabilizes sigma-model
strings, in the sense that a perturbation has a permanent effect, but this
effect remains small.

This discussion has assumed that the back-reaction on the space-time
metric is negligible, and it leaves open the question of whether the
situation changes when the back-reaction is included (or as
the coupling constant $\kappa$ mentioned earlier is turned on and increased).
It may be possible to extend moduli-space calculations such as those in
\cite{SS99,G01} so as to apply to this question.

%%%%%%%%%%%%%%%%%%%%%%%%%%%%%%%%%%%%%%%%%%%%%%%%%%%%%%%%%%%%%%%%%%

\section{Stability of Textures.}

In this section, we investigate the stability of textures in the O(4)
sigma model \cite{D87,T89,TS90,STPR91} and the O(3) sigma model
\cite{Luo92,SCP97,HL99}.  The natural tendency for textures
is to shrink and eventually decay, and the question is to what extent
cosmological expansion can enable them to avoid this fate.  Note that
O(3) textures are the same as closed sigma-model strings --- in these loops,
it is the circumference of the string which shrinks (and not its width).

Let us begin with the O(4) case.  In an open universe, there are no
comoving texture solutions.  In a closed Robertson-Walker universe
\begin{equation} \label{RWclosed}
 ds^2 = dt^2 - A(t)^2\,[d\xi^2+\sin^2\xi\,(d\theta^2+\sin^2\theta\,d\vp^2)],
\end{equation}
however, a comoving solution exists \cite{D87}.  Let us begin by
investigating O(4) textures in the background (\ref{RWclosed}),
using a $\l$-approximation.  Take the sigma-model field
$\Phi^a$ to have the form
\begin{equation} \label{O3ansatz}
  \Phi^a = (\sin{f}\sin{\theta}\cos{\vp},
    \sin{f}\sin{\theta}\sin{\vp}, \sin{f}\cos{\theta},\cos{f})\,,
\end{equation}
where $f=f(t,\xi)$.  This represents an O(3)-symmetric configuration.
The reduced Lagrangian (after doing the $\theta$ and $\vp$ integrations) is
\[
  L = 2\pi A\left[\sin^2{\xi}\,(A^2f_t^2-f_{\xi}^2)-2\sin^2{f}\right].
\]
For the $\l$-approximation, we take
$f(t,\xi)$ to be defined by $\tan{(f/2)}=\l(t)\,\tan{(\xi/2)}$.
This may be thought of as a stereographic projection from the spatial
3-sphere to $\RR^3$, followed by a dilation on $\RR^3$, and then by
an inverse stereographic projection to the target 3-sphere; the same
configurations have been used in the study of Skyrmions \cite{M87}.
Integrating over $\xi$ then leaves the effective Lagrangian
\begin{equation} \label{tex4lag}
  L = 12\pi^2\left[\frac{A^3\dot\l^2}{(1+\l)^4}
         - \frac{A\l}{(1+\l)^2}\right],
\end{equation}
and the corresponding equation of motion is
\begin{equation} \label{tex4eqn}
\ddot\l = \frac{2\dot\l^2}{1+\l} - \frac{3\dot A \dot\l}{A}
           - \frac{1-\l^2}{2A^2},
\end{equation}
which is the same as
\begin{equation} \label{tex4eqnB}
\frac{d}{dt}\left[\frac{A^3\dot\l}{(1+\l)^3}\right]
  = \frac{A(\l-1)}{2(\l+1)}.
\end{equation}

The comoving solution referred to earlier is $\l(t)\equiv1$, representing
a spatially-homogeneous texture.  If $A$ is constant
(no expansion), then this solution is unstable, as is
clear from (\ref{tex4eqn}) (this contradicts the claim for
stability made in \cite{D87}).  Note that the system is invariant
under $\l\mapsto\l^{-1}$; that collapse corresponds to either
$\l\to0$ or $\l\to\infty$; and that $\l$ (or $\l^{-1}$) may
be thought of as the ratio between the size of physical space
and the size of the texture.  So in particular for $\l$ close to zero,
the spatial curvature should play little role, and the results should
provide an indication of what happens in an open universe.

Let us now investigate the behaviour of solutions of (\ref{tex4eqnB})
in an expanding universe.  If the texture is highly localized
(say $0<\l\ll 1$), then (\ref{tex4eqnB}) can be approximately solved,
as follows.
Taking $A(t)\sim t^p$, we get $\l(t)\approx\a+\b t^{1-3p}+\g t^{2-2p}$,
where $\a$, $\b$ and $\g$ are constants.
It follows from this that if $p\leq1$, then $\l$ will inevitably become
zero (\ie\ the texture will collapse and decay) in finite time; but
if $p>1$, then (depending on initial conditions) it might survive
indefinitely.  In particular, in our current universe, small textures
certainly collapse.

What about the homogeneous texture, with $\l=1$?  Numerical solution
of (\ref{tex4eqn}) shows that the same is true.  Namely, start with
$\l=1$ and put in a (small) perturbation which acts to reduce $\l$. Then
for $p\leq1$, $\l(t)$ reaches 0 in finite time; while for $p>1$,
$\l(t)\to\l_{\infty}$ as $t\to\infty$, where $0<\l_{\infty}<1$.

We can confirm this picture by studying O(3)-symmetric textures in a
flat-space Robertson-Walker background.  The relevant reduced Lagrangian in
this case is
\[
  L = 2\pi A\left[r^2\,(A^2f_t^2-f_r^2)-2\sin^2{f}\right],
\]
where $f=f(t,r)$ with boundary conditions $f(t,0)=\pi$ and 
$f(t,\infty)=0$.  If $A(t)\equiv1$, the equations of motion admit
the well-known `collapsing' solution $f = 2\cot^{-1}{(-r/t)}$ for $t<0$
\cite{TS90}.  In the expanding case, the equations can be solved numerically,
and the result is as before: if $A(t)\sim t^p$ with $p>1$, then a small
perturbation causes textures to change their shape, but they do not
collapse.

For Hopf textures, \ie\ for the O(3) sigma model, one expects the results
to be very similar.  Imposing an ansatz analogous to (\ref{O3ansatz})
is no longer very natural, since Hopf structures cannot be O(3)-symmetric
(\cf\ \cite{HL99}). But there is a natural one-parameter family of
configurations labelled by $\l$ (\cf\ \cite{W99}), leading to an equation
similar to
(\ref{tex4eqnB}), and to the same conclusions.  The configuration with
$\l=1$ is an exact comoving spatially-homogeneous solution (in closed
Robertson-Walker space-time); it consists simply of the
standard Hopf map from the spatial $S^3$ to the target $S^2$.  So this
static solution is unstable, and all other Hopf textures
inevitably collapse, if the cosmological expansion is slower $A(t)\sim t$;
but during an era of faster expansion, they are stable in a generalized sense.

%%%%%%%%%%%%%%%%%%%%%%%%%%%%%%%%%%%%%%%%%%%%%%%%%%%%%%%%%%%%%%%%%

\end{document}